\documentclass[11pt,twoside]{article}
\usepackage{asp2014}
\usepackage[symbol]{footmisc}

\newcommand{\etal}{\textit{et al.}}

\aspSuppressVolSlug
\resetcounters

\begin{document}

\title{Six Dimensional Streaming Algorithm for Cluster Finding in N-Body Simulations }

\author{Aidan~Reilly$^1$, Nikita~Ivkin$^2$\footnotemark, Gerard~Lemson$^1$,  Vladimir~Braverman$^1$,  and Alexander~Szalay$^1$}
\affil{$^{1}$Johns Hopkins University , Baltimore, MD; \email{areilly8, glemson1, vbraver1, szalay@jhu.edu}}
\affil{$^{2}$ Amazon; \email{ivkin@amazon.com}}

\paperauthor{Gerard~Lemson}{}{ORCID}{Johns Hopkins University }{Author1 Department}{City}{State/Province}{Postal Code}{Country}
\paperauthor{Sample~Author2}{Author2Email@email.edu}{ORCID_Or_Blank}{Author2 Institution}{Author2 Department}{City}{State/Province}{Postal Code}{Country}



\titlefootnote{This work was done while the author was at Johns Hopkins University.} 
 
\begin{abstract}
Cosmological N-body simulations are crucial for understanding how the Universe evolves. Studying large-scale distributions of matter in these simulations and comparing them to observations usually involves detecting dense clusters of particles called ``halos,'' which are gravitationally bound and expected to form galaxies.  However, traditional cluster finders are computationally expensive and use massive amounts of memory. Recent work by Liu~\etal~(\citet{streaming}) showed the connection between cluster detection and memory-efficient streaming algorithms and presented a halo finder based on heavy hitter algorithm. Later, Ivkin~(\etal~\citet{scalable-streaming-tools}) improved the  scalability of suggested streaming halo finder with efficient GPU implementation. Both works map particles' positions onto a discrete grid, and therefore lose the rest of the information, such as their velocities. Therefore two halos travelling through each other are indistinguishable in positional space, while the velocity distribution of those halos can help to identify this process which is worth further studying. In this project we analyze data from the Millennium Simulation Project~(\citet{millennium}) to motivate the inclusion of the velocity into streaming method we introduce. We then demonstrate a use of suggested method, which allows one to find the same halos as before, while also detecting those which were indistinguishable in prior methods.
\end{abstract}

\section{Introduction}
A major effort in learning about the evolution of the universe involves the use of advanced computer simulations of the gravitational evolution of a system of particles. The goal is to compare the output of these simulations to observations or among each other, and consequently evaluate the assumptions and parameters used at the start of the simulation. The comparisons cannot be made particle to particle due to the large number of particles in each simulation and noisy nature of the data. Instead, researchers compare clusters of particles, ''halos'', the distribution of which can be more readily compared to the observer galaxy distribution. Informally, a halo is a region in the simulation space with a high mass concentration where we expect the particles to be gravitationally bound. Most methods for finding these halos in the simulation are computationally expensive and require loading all of the particles into memory~(\citet{comparison}). As a result, a simulation with $10^{12}$ particles~(\citet{millennium-xxl}) requires $12$ terabytes of memory in hardware. Recent results by \citet{streaming} and \citet{scalable-streaming-tools} present a method for halo finding which uses sublinear memory and runs in under an hour. It transforms the particle data and reduces the halo finding to finding frequent items in the stream. Finding frequent items (or ``heavy hitters'') is a well studied problem in the field of streaming algorithms. Both theoretically optimal and practically efficient algorithms were developed over the last two decades (\citet{CS,cormode2005improved,braverman2017bptree,misra1982finding}).
The main drawback of current streaming halo finders is the need to map particle positions onto a discrete grid of cells, and inserting the ID of the cell into a data stream.  This method forgoes all information apart from position, such as that given by the particles' velocities. As a simulation evolves, many distinct clusters will travel in close vicinity of each other, or even through each other, making them indistinguishable using only location information. However, these nearby clusters are not always gravitationally bound to one another and can often be distinguished by their velocities. Such close clusters with starkly different velocity distributions  signify interesting regions in the simulation that someone may want to identify quickly for further study. In this project we give a method for including particle velocities as part of the data transformation in a way that can be included in the streaming tools already developed with minimal changes. We implement a simplified version of the streaming tool here to show its efficacy in pinpointing such regions of interest.

\section{Contribution}

A data stream $S = q_1,q_2,...,q_n$ is a sequence of n objects  $q_i\in \mathcal{O} = \{o_1,...,o_m\}$. The $q_i$ are entities such as integers, edges in a graph, sets , images, web pages, etc. In the streaming model an algorithm makes a single (or at most a few) pass(es) over $S$ and uses sub-linear, $o(n+m)$ memory. This requires algorithms to be randomized and approximate, however for many practical purposes such an approximation is sufficient. The Count Sketch  algorithm (\cite{CS}) for example finds frequent items in the stream using only $O(\log nm)$ memory. \cite{streaming}  reduced the halo finding problem to finding heavy hitters (i.e. frequent items) in the density field, requiring gigabytes rather than terabytes of memory. 




 We work on data from a single snapshot of the Millennium Simulation Project (\citet{millennium}), a cosmological N-body simulation with $10^{10}$ particles. The simulation has a size of 500 Mpc/$h$. Each particle has its position and velocity information saved and we transform the data in a way that starts with the method used by \citet{streaming}, however it  includes the velocity. We choose the size of a position cell to be 1 Mpc/$h$ to match the size of a typical large halo, creating a cubic grid with side length $\ell = 500$ cells. 
Unlike the position of a particle, its velocity is in principle unbounded, and can  be negative as well as positive and thus must be handled differently. For ease of discretization, we use velocity as a scalar quantity, as we only ever analyze one direction of velocity at a time. When reading a particle with position (x,y,z) and velocity v:
 \small
 \begin{itemize}
    \item If v is less than -$s=2000$ km/s or greater than $s=2000$ km/s discard the particle, otherwise add $s$ to the velocity to shift all relevant particles to have positive velocities
    \item Determine it's cell by computing the function below and insert the ID into the data stream   
    $$\text{ID}(v, x, y, z) = v\cdot \frac{c}{2s} + (x + y\cdot\ell + z\cdot\ell^2)\cdot c$$ 
\end{itemize}
\normalsize

For simplicity we discard the particles with overall speed higher than 2000 km/s, the number of such particles is relatively small and should not influence the results. Note that we now have a total of $\ell^3\cdot c$ cells with unique IDs that we can pass in to a streaming algorithm. The $c$ velocity cells corresponding to each position cell in the 3 dimensional partition will give an approximate histogram of the velocities of the particles in that space. To find heavy hitters we use Count Sketch (\citet{CS}), the same algorithm was used by \cite{streaming} and \cite{scalable-streaming-tools}.

\par To motivate the use of the Millennium Simulation data, we performed a nearest neighbor search on the $10^5$ heaviest SUBFIND  (\citet{subfind}) halos obtained from the Millennium Database (\citet{MRDB}). Approximately 10$^4$ distinct halo pairs are within 1 Mpc/$h$ of each other, thus potentially mapping to the same position cell.
The mean velocity vectors for over 80 percent of these pairs are separated by >25 degrees, showing clear distinction in velocity space. High velocity dispersion could make this distinction disappear when looking at particle velocities rather than overall halo velocities. However, we found the number of cases where this happens to be relatively small.
The distinction can be seen when looking at velocity in only one direction at a time, motivating our design. We run 4 separate sketches in parallel and then take a union of their results. The pipeline is as follows:
\small
\begin{itemize}
    \item Partition the space using 3 dimensions, partition the space three times using 4 dimensions (once for each velocity direction), and run Count-Sketch on all 4 partitions in parallel
    \item For all heavy cells found in the 3 dimensional run: search the velocity cells corresponding to that space in the results of the other 3 runs
    \item If any of the 4 dimensional runs show two distinct peaks in the approximate velocity histograms generated for that position, identify this area as having two clusters and get the approximate particle count in each one
\end{itemize}
\normalsize
Continuing to run the 3 dimensional version assures that we do not miss any halos that we would have picked before including velocity. We then include each dimension of velocity separately  because it facilitates discretization and the results of our analysis show this to be sufficient. 
\section{Conclusions}
\articlefigure[height=3cm,width=12cm]{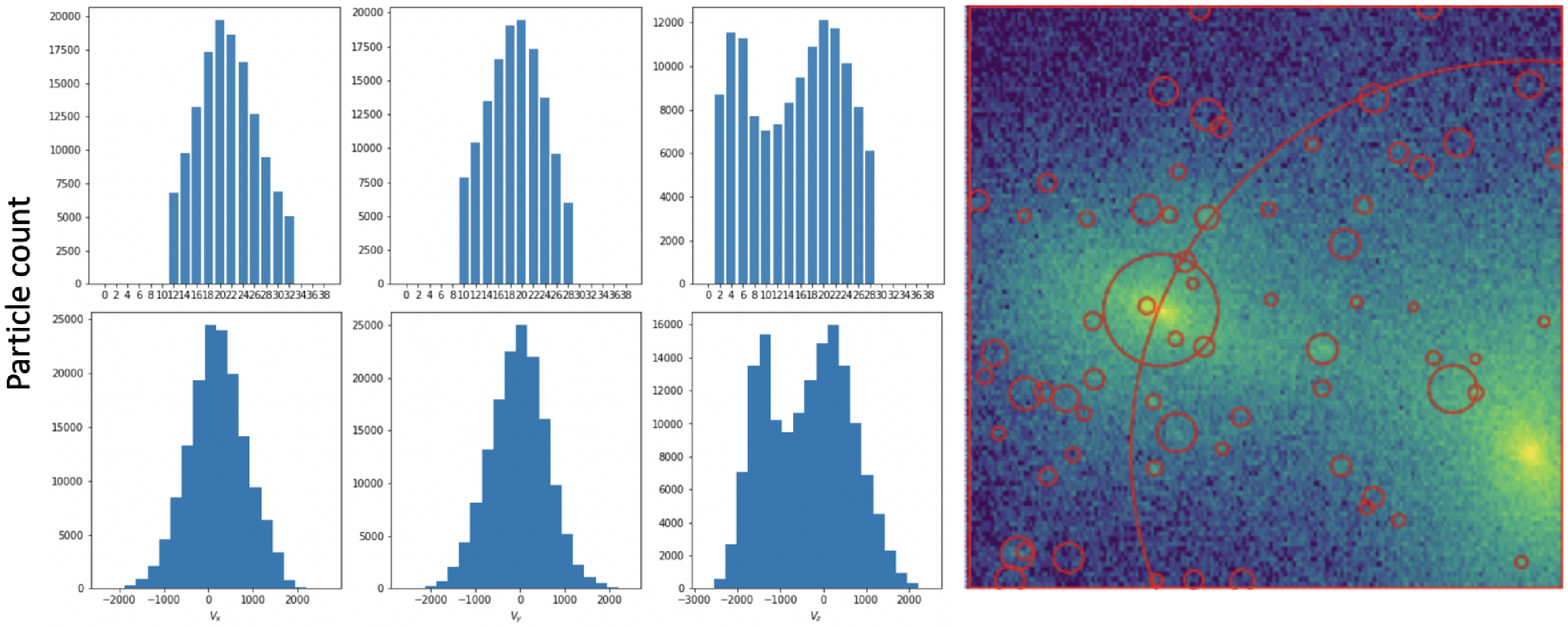}{Fig1}{(Top-left) Approximated velocity distribution of a single position cell detected as a region of interest. (Bottom-left) Exact velocities at same cell. (Right) Density map of the particles in the same cell, showing we have found a clear double halo.
}
We compare our results to the output of an exact heavy hitter under the same data transformation as it gives an exact histogram of velocities at each position. With sketching we can capture up to 91 percent of the 267 regions of interest tagged by the exact algorithm. Figure 1 shows an example of one such region, and Figure 2 shows how the memory used is proportional to the fraction of interesting regions found.  If one velocity cell at a certain position does not show up in the top 10$^5$ heaviest cells while two surrounding it do, it would falsely look like two peaks to our search algorithm, and so we exclude any interesting regions where the middle of two peaks drops all the way to zero. By doing this the we raise our accuracy by 30 percent. It is likely, however, that this method causes us to miss some regions, and a better method for determining two peaks needs to be investigated further.  We give the approximated particle count per peak found in interesting regions so that a user can focus their analyses based on these values. Now that we have solid evidence for the usefulness of this method, we plan to incorporate it into the full scalable streaming tool developed in \citet{streaming}.
\articlefigure[height=4cm,width=8cm]{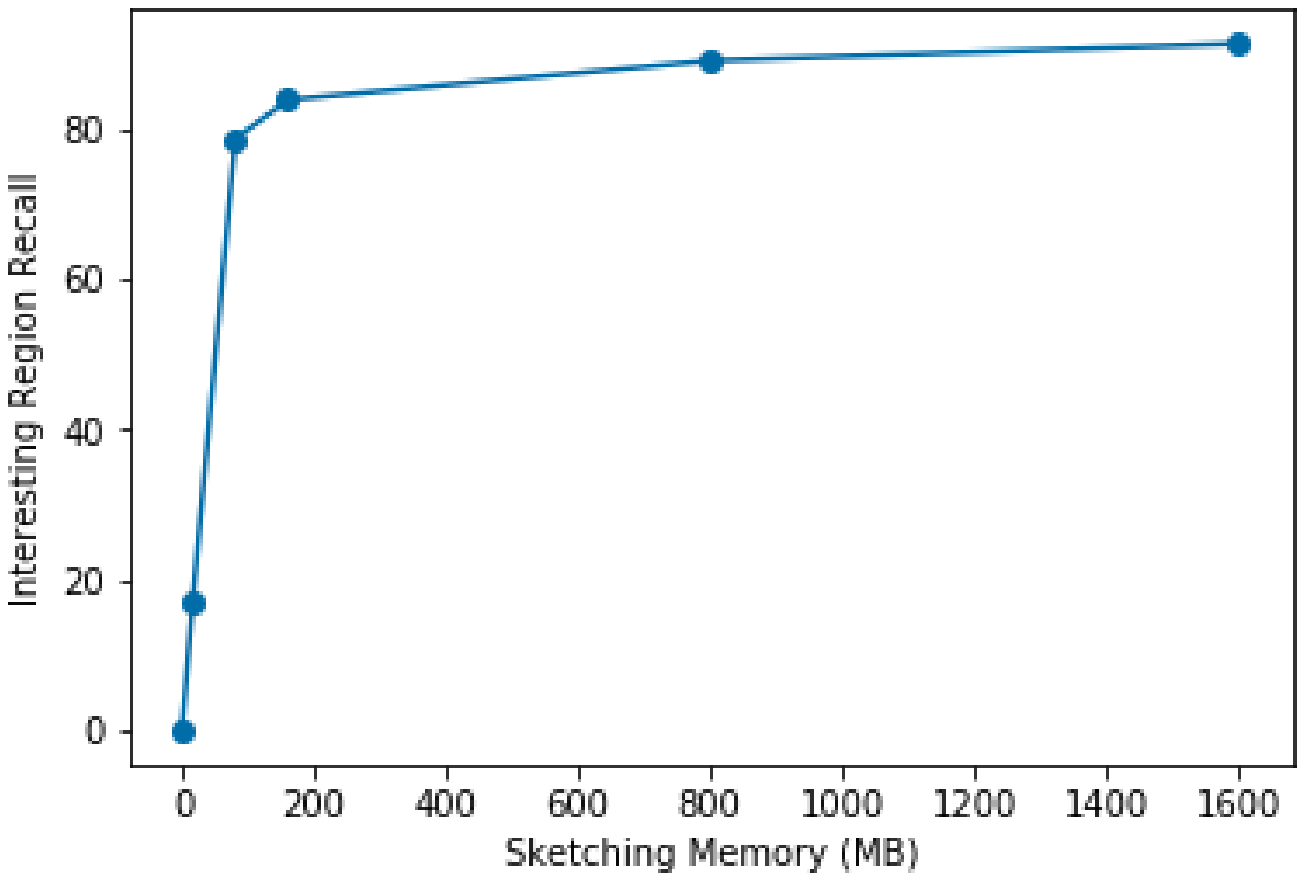}{Fig2}{Region of interest percent recall vs. Memory of sketching data structure }

\bibliography{P3-4}

\begin{thebibliography}{}
\expandafter\ifx\csname natexlab\endcsname\relax\def\natexlab#1{#1}\fi
\expandafter\ifx\csname url\endcsname\relax
  \def\url#1{\texttt{#1}}\fi
\expandafter\ifx\csname urlprefix\endcsname\relax\def\urlprefix{URL }\fi
\providecommand{\eprint}[2][]{\url{#2}}

\bibitem[{Angulo et~al.(2012)Angulo, Springel, White, Jenkins, Baugh, \&
  Frenk}]{millennium-xxl}
Angulo, R., Springel, V., White, S., Jenkins, A., Baugh, C., \& Frenk, C. 2012,
  Monthly Notices of the Royal Astronomical Society, 426, 2046

\bibitem[{Braverman et~al.(2017)Braverman, Chestnut, Ivkin, Nelson, Wang, \&
  Woodruff}]{braverman2017bptree}
Braverman, V., Chestnut, S.~R., Ivkin, N., Nelson, J., Wang, Z., \& Woodruff,
  D.~P. 2017, in Proceedings of the 36th ACM SIGMOD-SIGACT-SIGAI Symposium on
  Principles of Database Systems (ACM), 361

\bibitem[{Charikar et~al.(2002)Charikar, Chen, \& Farach-Colton}]{CS}
Charikar, M., Chen, K., \& Farach-Colton, M. 2002, in International Colloquium
  on Automata, Languages, and Programming (Springer), 693

\bibitem[{Cormode \& Muthukrishnan(2005)}]{cormode2005improved}
Cormode, G., \& Muthukrishnan, S. 2005, Journal of Algorithms, 55, 58

\bibitem[{Ivkin et~al.(2018)Ivkin, Liu, Yang, Kumar, Lemson, Neyrinck, Szalay,
  Braverman, \& Budavari}]{scalable-streaming-tools}
Ivkin, N., Liu, Z., Yang, L.~F., Kumar, S.~S., Lemson, G., Neyrinck, M.,
  Szalay, A.~S., Braverman, V., \& Budavari, T. 2018, Astronomy and computing,
  23, 166

\bibitem[{Knebe et~al.(2011)Knebe, Knollmann, Muldrew, Pearce, Aragon-Calvo,
  Ascasibar, Behroozi, Ceverino, Colombi, Diemand et~al.}]{comparison}
Knebe, A., Knollmann, S.~R., Muldrew, S.~I., Pearce, F.~R., Aragon-Calvo,
  M.~A., Ascasibar, Y., Behroozi, P.~S., Ceverino, D., Colombi, S., Diemand,
  J., et~al. 2011, Monthly Notices of the Royal Astronomical Society, 415, 2293

\bibitem[{Lemson \& the Virgo~Consortium(2006)}]{MRDB}
Lemson, G., \& the Virgo~Consortium 2006, astro-ph/0608019

\bibitem[{Liu et~al.(2015)Liu, Ivkin, Yang, Neyrinck, Lemson, Szalay,
  Braverman, Budavari, Burns, \& Wang}]{streaming}
Liu, Z., Ivkin, N., Yang, L., Neyrinck, M., Lemson, G., Szalay, A., Braverman,
  V., Budavari, T., Burns, R., \& Wang, X. 2015, in 2015 IEEE 11th
  International Conference on e-Science (IEEE), 342

\bibitem[{Misra \& Gries(1982)}]{misra1982finding}
Misra, J., \& Gries, D. 1982, Science of computer programming, 2, 143

\bibitem[{Springel et~al.(2005)Springel, White, Jenkins, Frenk, Yoshida, Gao,
  Navarro, Thacker, Croton, Helly et~al.}]{millennium}
Springel, V., White, S.~D., Jenkins, A., Frenk, C.~S., Yoshida, N., Gao, L.,
  Navarro, J., Thacker, R., Croton, D., Helly, J., et~al. 2005, nature, 435,
  629

\bibitem[{{Springel} et~al.(2001){Springel}, {White}, {Tormen}, \&
  {Kauffmann}}]{subfind}
{Springel}, V., {White}, S.~D.~M., {Tormen}, G., \& {Kauffmann}, G. 2001,
  \mnras, 328, 726. \eprint{astro-ph/0012055}

\end{thebibliography}


\end{document}